\documentclass[manuscript]{aastex63}

\usepackage{amsmath,amstext}
\usepackage[T1]{fontenc}
\usepackage[figure,figure*]{hypcap}
\usepackage{yfonts}
\usepackage{bbold}

\usepackage{algorithm}
\usepackage[noend]{algpseudocode}

\usepackage{comment}

 \turnoffedit 
 
\newcommand{\gsim}{\lower.7ex\hbox{$\;\stackrel{\textstyle>}{\sim}\;$}}
\newcommand{\lsim}{\lower.7ex\hbox{$\;\stackrel{\textstyle<}{\sim}\;$}}

\usepackage{listings} 
\usepackage{color}

\definecolor{codegreen}{rgb}{0,0.6,0}
\definecolor{codegray}{rgb}{0.5,0.5,0.5}
\definecolor{codepurple}{rgb}{0.58,0,0.82}
\definecolor{backcolour}{rgb}{0.95,0.95,0.92}

\lstdefinestyle{mystyle}{
    backgroundcolor=\color{backcolour},   
    commentstyle=\color{codegreen},
    keywordstyle=\color{magenta},
    numberstyle=\tiny\color{codegray},
    stringstyle=\color{codepurple},
    basicstyle=\footnotesize,
    breakatwhitespace=false,         
    breaklines=true,                 
    captionpos=b,                    
    keepspaces=true,                 
    numbers=left,                    
    numbersep=5pt,                  
    showspaces=false,                
    showstringspaces=false,
    showtabs=false,                  
    tabsize=2
}

\shortauthors{Holman et al.}

\submitjournal{PSJ, 27 January 2023}
\revised{28 March 2023}

\begin{document}

\title{ASSIST: An Ephemeris-Quality Test Particle Integrator}

\author[0000-0002-1139-4880]{Matthew~J.~Holman}
\affiliation{Center for Astrophysics | Harvard \& Smithsonian, 60 Garden St.,  Cambridge, MA 02138, USA}
\correspondingauthor{Matthew~J.~Holman}
\email{mholman@cfa.harvard.edu}

\author[0000-0003-4173-4978]{Arya Akmal} 
\affiliation{Montgomery College, Rockville, MD 20850, USA}
\affiliation{Center for Astrophysics | Harvard \& Smithsonian, 60 Garden St., Cambridge, MA 02138, USA}

\author[0000-0003-0774-884X]{Davide Farnocchia} 
\affiliation{Jet Propulsion Laboratory, California Institute of Technology, 4800 Oak Grove Dr., Pasadena, CA 91109, USA}

\author[0000-0003-1927-731X]{Hanno Rein} 
\affiliation{Department of Astronomy and Astrophysics, University of Toronto, Toronto, Ontario, M5S 3H4, Canada}

\author[0000-0001-5133-6303]{Matthew~J.~Payne} 
\affiliation{Center for Astrophysics | Harvard \& Smithsonian, 60 Garden St., MS 51, Cambridge, MA 02138, USA}

\author[0000-0002-0439-9341]{Robert Weryk} 
\affiliation{Department of Physics and Astronomy, The University of Western Ontario, 1151 Richmond Street, London ON N6A 3K7, Canada}

\author[0000-0002-9908-8705]{Daniel Tamayo} 
\affiliation{Department of Physics,
Harvey Mudd College,
301 Platt Blvd.,
Claremont, CA 91711, USA}

\author[0000-0001-7648-0926]{David M. Hernandez} 
\affiliation{Center for Astrophysics | Harvard \& Smithsonian, 60 Garden St., MS 51, Cambridge, MA 02138, USA}
\affiliation{Yale University, 52 Hillhouse, New Haven, CT 06511, USA}


\begin{abstract}
We introduce ASSIST, a software package for ephemeris-quality integrations of test particles.  ASSIST is an extension of the REBOUND framework  
and makes use of its IAS15 integrator
to integrate test particle trajectories in the field of the Sun, Moon, planets, and 16 massive asteroids, with the positions of the masses coming from the JPL DE441 ephemeris and its associated asteroid perturber file. The package incorporates the most significant  gravitational harmonics and general relativistic corrections. ASSIST also  accounts for position- and velocity-dependent non-gravitational effects.  
The first order variational equations are included for all terms to support orbit fitting and covariance mapping. This new framework is meant to provide an open-source package written in a modern language to enable high-precision orbital analysis and science by the small body community.
ASSIST is open source, freely distributed under the GNU General Public license, version 3. 

\end{abstract}


\section{Introduction}
\label{SECN:INTRO}

It is well known that the $n$-body equations of motion involving gravitational interactions among more than two bodies \edit2{ are not classically integrable}~\citep{Murray.2000}. 
Practical applications thus require perturbative or numerical treatment. Numerical approaches have become increasingly accurate since the advent of fast electronic computers, beginning in the second half of the twentieth century. The concomitant development of modern algorithms for this purpose has advanced significantly over the past few decades.

The introduction of symplectic integrators for the $n$-body problem contributed to a resurgence of interest and research in {\it long-term} solar system dynamics~\citep{Gladman.1991,Wisdom.1991}.  For numerical integrations of systems over billions of years it is particularly important to capture key properties, such as conservation of energy and angular momentum, secular and mean motion resonances, and intrinsic dynamical chaos.  
However, with the availability of extremely precise observations, such as radar ranging \citep{Ostro2002}, stellar occultations \citep{Ferreira2022}, and space-based astrometry \citep{Tanga.2022}, the study of {\it short-term} dynamics has becoming increasing relevant.  That is the focus of this paper.

The specific problem we consider is that of a massless \emph{test particle} moving in the gravitational field of the Sun, planets, Moon, and a set of massive asteroids, with the pre-computed, time-dependent  positions and velocities of these massive bodies provided by reliable ephemerides. 
For this, we develop ASSIST, a software package for ephemeris-quality integrations of test particles.  ASSIST is an extension of the REBOUND framework~\citep{Rein.2012} 
and makes use of its IAS15 integrator~\citep{Rein.2015}. The emphasis is accuracy, with a model that encompasses all the relevant physical effects.  Possible applications include accurately fitting orbits and detailed tests of gravity itself. 

The standard of reference for solar system small body ephemeris computations is
the Horizons service,\footnote{\url{https://ssd.jpl.nasa.gov/horizons/}} developed and maintained by the Solar System Dynamics Group at the Jet Propulsion Laboratory (JPL). 
Although the underlying source code is not public, the Horizons system is publicly accessible via a web interface, as well as through an API.
The JPL small-body orbit software is an ideal basis of comparison since its accuracy has been thoroughly tested by observations as well as space missions~\citep[e.g.,][]{Farnocchia.2021}. 

It should be noted that a number of software packages have been developed and are available specifically for integrating and fitting orbits over observational time scales.  These include {\it OrbFit} from the University of Pisa's Celestial Mechanics Group~\citep{Orbfit.2011}, {\it Find\_Orb} from Project Pluto\footnote{\url{https://www.projectpluto.com/find\_orb.htm}}, {\it OpenOrb} from the University of Helsinki~\citep{Granvik.2009}, and the package developed by \citet{Bernstein.2000}, also named {\it orbfit}, for fitting the orbits of transneptunian objects.  Each, as well as others, is used effectively by members of the small body community.

This raises an obvious question: why develop another such package?  Our motivations for developing a new integrator include using state of the art numerical techniques, using a more current language (C99), providing Python wrappers so that the tools are accessible to a wider audience, and transitioning away from file-based processing to enhance computational speed. \edit2{(We note that {\it OpenOrb} already addresses the last two objectives.)}

In addition to matching the integrated output of Horizons, we also have technical goals for the integrator that are driven by the precision of current and future data sets.
\edit3{Gaia} can deliver sub-milli-arcsecond (mas) astrometry for bright asteroids~\citep{Tanga.2022}.  Gaussian process regression, applied to Dark Energy Camera astrometry of the TNO Eris, reached residuals of 5~mas~\citep{Fortino.2021AJ}.  \edit2{Astrometric precision of 1-2~mas should be achievable} with the Rubin Observatory's Legacy Survey of Space and Time (LSST).  Thus, it is essential that the precision of the numerical integration tools used to model such objects exceeds that of the data.  We adopt as a target that the numerical errors for a main belt asteroid result in an astrometric uncertainty of  no more than 1~milli-arcsecond over a ten year integration.  The JPL small body code meets these goals, but it is not available to the community other than through web interfaces.

The remainder of this paper is organized as follows.
In \S\ref{SECN:INTEGRATOR}, we describe the overall integrator.
In \S\ref{SECN:EQUATIONS}, we present the equations of motion.  
In \S\ref{SECN:TESTS}, we present tests of the integrator and its components.
In \S\ref{SECN:CODE}, we describe the Python wrapper.
Finally, in \S\ref{SECN:DISCUSS}, we discuss the implications of our results and describe possible future work.

\section{REBOUND IAS15 Integrator}
\label{SECN:INTEGRATOR}

ASSIST is built in the framework of the REBOUND package~\citep{Rein.2012}. REBOUND is a collection of numerical integrators, with associated tools, designed to carry out investigations of solar system dynamics.  It provides a consistent, well-documented environment.  The code is written in C99 and is available on Github
\footnote{\url{https://github.com/hannorein/rebound}}.

Although other integrators included in REBOUND could be used with the equations of motion described in Sec.~\ref{SECN:EQUATIONS}, we focus on the IAS15 integrator~\citep{Rein.2015}.  IAS15 is a predictor-corrector method, based on the RADAU integrator~\citep{Everhart.1985}.  However, IAS15 includes compensated summation~\citep{Kahan.1965}. 
It is highly efficient, and for some problems and specific phase space variables, IAS15's error grows as the square root of the number of steps, which is optimal~\citep{Brouwer.1937}.
Although this is not fully appreciated, even by many dynamicists, the RADAU and IAS15 integrators are general purpose tools that happen to be very well suited to integrating equations of motion that have very smooth solutions, such as the $n$-body problem~\citep{Everhart.1985}. 
Therefore, it is straightforward to include additional terms in the equations of motion.  

The REBOUNDx extension to REBOUND provides a convenient, general framework for the addition of such terms to the equations of motion and is the natural choice for most such work~\citep{Tamayo.2020}.  However, although the mechanism of extending REBOUND is similar to that of REBOUNDx, ASSIST is a highly specialized package that is far less flexible by design and necessity. 
Because the positions and velocities of the perturbers come from the DE441 ephemeris~\citep{Park.2021}, the coordinate frame, time scale, masses, and other physical constants are fixed and the test particles must abide by those.  We chose to develop ASSIST as a distinct package to enforce self-consistency of those units, as well as to limit the dependencies of ASSIST to only the REBOUND package.

\edit3{Although ASSIST imposes this structure, the rest of the process is inherited from REBOUND and will be familiar to its users.  As is demonstrated in the examples, one initiates a REBOUND simulation, loads the ephemeris file, attaches the equations of motion from ASSIST, and adds test particles to the simulation (the planets and other perturbers are provided by ASSIST.  At that point, the routines for carrying and managing the integrations are essentially the same.}

\section{Equations of Motion}
\label{SECN:EQUATIONS}

We integrate the orbits of small bodies as massless test particles.  That is, they are subject to the gravitational influence of the massive bodies in the system, and may also be subject to non-gravitational effects, but they do not themselves affect the trajectories of their perturbers.
The time scale used is barycentric dynamical time (TDB)\footnote{\url{https://www.iau.org/static/resolutions/IAU2006_Resol3.pdf}}.
The coordinates are barycentric and in the International Celestial Reference Frame (ICRF)~\citep{Charlot.2020}.

\subsection{Perturbers}
\label{SSECN:PERTURBERS}

 In contrast to the usual IAS15 $n$-body integrator, which integrates the orbits of systems of fully interacting massive bodies, along with the orbits of test particles moving in the field of massive bodies, in our integrator, the positions of the perturbing objects are known as a function of time.
 We include the direct Newtonian acceleration from the Sun, planets, Moon, Pluto, and 16 massive asteroids. 
 
 The positions of the Sun, Moon, planets, and Pluto are given by JPL's DE441 ephemeris~\citep{Park.2021}.  
 These are determined by evaluating Chebyshev polynomials tabulated in binary SPK format~\citep{Acton.2018}.  
 These provide the polynomial coefficients in time segments that span the years -13200 to 17191.
 The extracted positions are barycentric, in the ICRF equatorial frame.  As noted earlier, we adopt the same frame for all internal calculations.  This avoids unnecessary coordinate transformations and simplifies the equations of motion.
 
 There is a corresponding file for a set of 16 massive asteroids, also in SPK format \citep{Farnocchia2021sb441}.\footnote{\url{ftp://ssd.jpl.nasa.gov/pub/eph/small_bodies/asteroids_de441/SB441_IOM392R-21-005_perturbers.pdf}}
 The positions of the massive asteroids are also in an equatorial frame but are heliocentric.  We convert these internally to barycentric, using the position of the Sun from the DE441 ephemeris, for consistency.

We developed library routines to access the SPK-format files. These routines make use of highly optimized ``memory mapping'' for  higher performance, while allowing thread-safe concurrent access with transparent disk caching.   The independent variable in the ephemeris routines is the number of days (TDB) relative to a specified reference date.  This avoids the loss of precision in representing time that would come with a large number of unchanging leading digits in the full Julian date.  It can be important to retain this precision when small time steps are used, such as during close encounters.  We adopt J2000.0 (JD 2451545.0/2000-01-01.5 TDB) as the reference date.

\subsection{Relativistic Corrections}
\label{SSECN:GR}

We implement three models to account for the effects of general relativity.  These can be selected by the user. The first includes only the central potential for the Sun that matches the apsidal precession rate from GR~\citep{Nobili.1986}.  The second includes the effects from the Sun only with the following expression~\citep{Damour.1985,Farnocchia.2015}:  
\begin{equation}
    \mathbf{a}_{REL} = \frac{GM_\odot}{c^2 |\mathbf{r}|^3}\left[\left(\frac{4GM_\odot}{|\mathbf{r}|}-|\mathbf{v}|^2\right)\mathbf{r} + 4\left(\mathbf{r}\cdot\mathbf{v}\right)\mathbf{v}\right],
\end{equation}
where $\mathbf{r}$ and $\mathbf{v}$ in this expression are the heliocentric position and velocity of the test particle, and $\mathbf{a}_{REL}$ is the heliocentric acceleration due to solar GR with this simplified model. These two models are faster but less accurate than the Parameterized Post-Newtonian (PPN, also known as Einstein-Infeld-Hoffman) formulation for the Sun and planets, assuming the body whose orbit is being integrated is a test particle~\citep{Moyer2003}.  The equations of motion for point particles in the PPN formulation, reproduced for reference, are as follows
\begin{eqnarray}
 \label{eq:ppn}
    \ddot{\mathbf{r}}_i &=& \sum_{j\neq i} \frac{\mu_j\left(\mathbf{r}_j - \mathbf{r}_i\right)}{r_{ij}^3}
    \left\{1 - \frac{2(\beta + \gamma)}{c^2} \sum_{l\neq i} \frac{\mu_l}{r_{il}} -\frac{2\beta-1}{c^2}\sum_{k\neq j}\frac{\mu_k}{r_{jk}}
     + \gamma \left(\frac{v_i}{c}\right)^2
     + (1+\gamma) \left(\frac{v_j}{c}\right)^2\right.\\\nonumber
     && \left. -\frac{2 (1+\gamma)}{c^2} \dot{\mathbf{r}}_i\cdot \dot{\mathbf{r}}_j
   -\frac{3}{2c^2}\left[\frac{(\mathbf{r}_i-\mathbf{r}_j)\cdot \dot{\mathbf{r}}_j}{r_{ij}}\right]^2
   + \frac{1}{2c^2}\left( \mathbf{r}_j-\mathbf{r}_i\right)\cdot \ddot{\mathbf{r}}_j
   \right\}\\\nonumber
   && + \frac{1}{c^2}\sum_{j\neq i} \frac{\mu_j}{r_{ij}^3}\left\{\left[ \mathbf{r}_i - \mathbf{r}_j\right]\cdot\left[ (2+2\gamma)\dot{\mathbf{r}}_i-(1+2\gamma)\dot{\mathbf{r}}_j\right]\right\}  
   \left(\dot{\mathbf{r}}_i - \dot{\mathbf{r}}_j\right)
   \\\nonumber
   && + \frac{3+4\gamma}{2c^2}\sum_{j\neq i} \frac{\mu_j\ddot{\mathbf{r}}_j}{r_{ij}}
      + \sum_{m=1}^{16}\frac{\mu_m (\mathbf{r}_m -\mathbf{r}_i)}{r_{im}^3},
\end{eqnarray}
which is accurate to $O(v^2/c^2)$.
The gravitational parameter for body $j$ is given by $\mu_j = G M_j$, where $G$ is the gravitational constant and $M_j$ is the gravitational mass of the body. For this expression, the position vectors $\mathbf{r}_i$ of the test particle and massive bodies are barycentric.  The Euclidean distance between bodies $i$ and $j$ is given by $r_{ij}$.  The speed of light is given by $c$.  
We assume $\beta=\gamma=1$, as is predicted by general relativity~\citep{Will.1971}, but these values can be easily changed by the user. The velocities are given by $v_i = |\dot{\mathbf{r}}_i|$.   The final summation over asteroid indices $m$ in equation~\ref{eq:ppn} reflects the fact that the GR corrections due to the massive asteroids are insignificant and can be ignored (the index $j$ does not include asteroids.)  
 The PPN model is used by the JPL small body integrator. (The Horizons service also uses the PPN model but restricts the outer loop in equation~\ref{eq:ppn} to the Sun.)

We take one of two approaches for the accelerations $\ddot{\mathbf{r}}_j$ on the right hand side of equation~\ref{eq:ppn}.
One solution is to calculate the acceleration of the planets; the Newtonian acceleration from the other bodies, excluding the massive asteroids, is sufficiently accurate, given that the PPN formulation is $O(v^2/c^2)$.  This is the approach recommended in \citet{Moyer2003}. 
Another solution is to calculate the acceleration from the JPL ephemeris by taking the next derivative of the Chebyshev polynomials (described in \ref{SSECN:PERTURBERS}). 
We note that the DE441 ephemeris files are Chebyshev fits to position with the velocity determined from its derivative; the interpolated accelerations are not guaranteed to be accurate.
However, our integrated results have matched those of the JPL  small-body code.  \edit2{Using the interpolated accelerations is the default behavior in ASSIST, but code for calculating the acceleration can be swapped in.}

\subsection{Gravitational Harmonics}
\label{SSECN: HARMON}

An accurate description of the gravitational field should account for the true shape of massive bodies~\citep{Battin.1987}.  
 We start with the oblateness of the Sun, which is well-described by the $J_2$ zonal harmonic; higher order zonal harmonics are negligible. The solar $J_2$ coefficient is $2.2\times10^{-7}$~\citep{Park.2021}. 

Additional higher order gravitational harmonics of the Earth are necessary for an accurate treatment of near Earth orbits. These are especially important during  close approaches.  
The Earth's $J_2$, $J_3$, $J_4$ and $J_5$ zonal harmonic coefficients are 
$1.08\times10^{-3}$, $-2.53\times10^{-6}$, $-1.62\times10^{-6}$, $-2.28\times10^{-7}$, respectively.  The full precision of these values are available in the DE441 headers~\citep{Park.2021}. Our current implementation truncates the zonal harmonic potential expansion for the Earth at $J_4$, and does not include the $J_5$ term. 

The mass distribution of the Earth has non-negligible azimuthal asymmetry.  
However, we do not account for axial asymmetry effects and include no non-zonal harmonic terms. 
These terms may be significant for near-Earth orbits and can be implemented in future work.  Such terms are included by JPL on an {\it ad hoc} basis when needed for specific orbits. For example, \citet{Naidu.2021} used the full $4\times4$ gravity field of the Earth, including non-zonal harmonics, to accurately model the trajectory of 2020 CD3, a temporarily captured ``mini-moon'' of the Earth.

\subsection{Non-Gravitational Effects}
\label{SSECN:NONGRAV}

The orbits of bodies moving under the influence of Newtonian gravity and relativistic corrections may also be affected by non-gravitational effects that depend on the intrinsic properties of the body.  For asteroids, these include solar radiation pressure~\citep{Vokrouhlicky.2000} and the Yarkovsky effect~\citep{Yokrouhlicky.2015}.  For comets, non-gravitational forces can be induced by outgassing from the body.  

A detailed, \emph{ab initio} model of such effects is not practical, as it would depend on typically unknown details of object shape, composition, and surface properties.
We include these effects with the semi-empirical, parameterized model of \citet{Marsden.1973}. 
These perturbations are reasonably expected to depend on the position and velocity of the object, in addition to its intrinsic properties. The parameterized acceleration is given by:
\begin{equation}
    \ddot{\mathbf{r}} = g(r)\left[
     A_1 \frac{\mathbf{r}}{r}
    +A_2 \frac{\mathbf{(r\times \dot{r})\times r}}{|\mathbf{(r\times \dot{r})\times r}|}
    +A_3 \frac{\mathbf{r\times \dot{r}}}{|\mathbf{r\times \dot{r}}|},
    \right]
    \label{eq:non-gravs}
\end{equation}
where  $A_1, A_2\  \& \ A_3$ are tunable parameters, with units of acceleration, that incorporate the intrinsic properties of the object. The three terms account for accelerations in the radial, transverse - in the direction of motion and normal (to the orbital plane) directions, respectively.  This assumes that the prefactor $g(r)$ of eqn.~\ref{eq:non-gravs} is general, as described below.

The prefactor models the dependence of the absorbed solar radiation (for asteroids) and degree of outgassing (for comets) on distance from the Sun. 
It is given by:
\begin{equation}
\label{eq:gofr}
    g(r) = \alpha\, (r/r_0)^{-m} 
    \left[1+(r/r_0)^n \right]^{-k}.
\end{equation}
The parameters $\alpha$, $r_0$, $m$, $n$ and $k$ are  specific to each  body being modelled.  The function $g(r)$ is unitless.
The value of $r_0$ is a normalizing distance.
The constants $m$, $n$, and $k$ model the decay of the effect of non-gravitational acceleration at distance $r$, and $\alpha$ is a normalization constant~\citep{Yeomans.2004}.  The constant $\alpha$ is set so that $g(r)=1$ for $r=1$~AU. For asteroids $g(r)$ reduces to 
$\left(1~\mathrm{AU}/r\right)^2$~\citep{Farnocchia.2015}.  For comets we set $g(r)$ to the standard expression in \citet{Marsden.1973}, in which case $r_0$ indicates the distance from the Sun where the solar insolation primarily contributes to the sublimation of ices. 

 \subsection{Variational Equations}
\label{SSECN:VARIS}

 In addition to the equations of motion, we implement the first order variational equations~\citep{Rein.2016}.   
  These describe the evolution of the analytic partial derivatives of each component of a test particle's acceleration with respect to each of the components of its initial conditions $\mathbf{r}$ and $\mathbf{\dot{r}}$, as well as its non-gravitational parameters $A_1, A_2, A_3$.  The variational equations are essentially the equations of motion for the phase space difference between neighboring trajectories, for small initial displacements.  These allow one to estimate the Lyapunov exponents of a given trajectory, as well as support orbit fitting and covariance mapping~\citep{Rein.2016}.  The alternative is to compute these by finite differences by integrating additional test particles with slightly offset initial conditions, but that process can be slower, more subject to numerical error and possibly limited by the extent of phase space.   We include terms for the variational equations for all the implemented effects (Newtonian gravity; Earth $J_2$, $J_3$ and $J_4$; solar $J_2$; the~\citet{Damour.1985} GR treatment; the Einstein-Infeld-Hoffman GR treatment; and the \citet{Marsden.1973} formulation of non-gravitational forces).
The first order variational equations are integrated along with the equations of motion of the corresponding test particle. The form of these equations is available in the code.\footnote{\url{https://github.com/matthewholman/assist/blob/main/src/forces.c}}

\section{Tests}
\label{SECN:TESTS}

We carry out a number of tests to demonstrate the correctness of the equations of motion and the integrator.  

\subsection{Force model terms}

To verify the equations of motion, we completed a term-by-term comparison of the various contributions to the acceleration of a test particle to the corresponding terms from the JPL small-body code.   Rather than integrating, we simply evaluate the acceleration of the test particle, asteroid (3666) Holman, at a specific epoch (JD 2458849.5 TDB, 2020 January 1).  This is a fine-grain test that verifies that the physical constants, the values of the positions and velocities of the perturbing bodies from the ephemeris, and  the algebraic terms in the equations of motion are identical to machine precision. Our code, as well as JPL's, is designed to work well in double precision for computational speed. 
As the individual terms span orders of magnitude, a term by term comparison is more stringent than comparing the summed accelerations.
Specifically, this comparison includes the gravitational constants ($\mathrm{GM}$ values); the components of the relative vector between the test particle and the Sun, planets, Moon, and massive asteroids; the direct Newtonian accelerations from those bodies; the complete PPN formulation of GR; the simplified formulation of GR; the $J_2$, $J_3$, and $J_4$ gravitational harmonics from the Earth, the solar $J_2$ gravitational harmonic, and the parameterization of non-gravitational forces presented earlier.  We find that all terms in our code agree with those from the JPL small-bodies integrator, differing only in the least significant digit. 

In order to test the PPN formulation of GR, which involves a large number of terms, we carried out a more detailed comparison of each contribution in equation~\ref{eq:ppn}.
This suite of comparisons verifies that our coding of the various contributions in the equations of motion is consistent with that of the corresponding small body code from JPL. %

\subsection{Variational equations} 

\begin{figure}[!tbph]
  \centering
    \includegraphics[width=\textwidth]{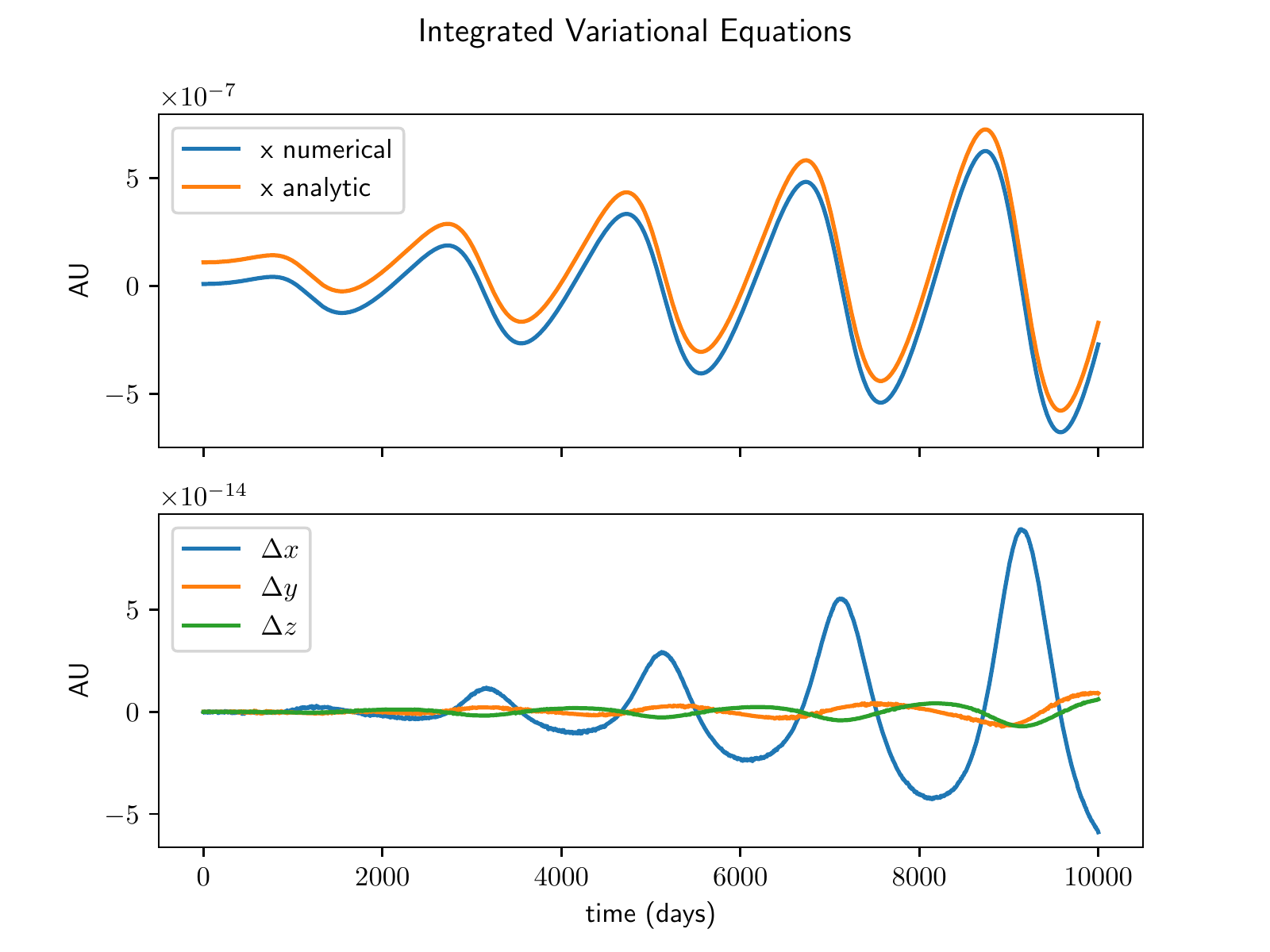}
      \caption{Top panel: Numerically and analytically calculated variational equations for the x-component of asteroid (3666) Holman.  The two curves have been intentionally offset by $10^{-7}$~AU to make them distinguishable on this scale. Bottom panel: Difference between the numerically and analytically calculated results, for all cartesian components. The magnitude of the difference grows quadratically in time, as expected. }
  	\label{FIG:vary}
\end{figure}

We carry out a detailed check of the variational equations with two approaches.
\edit2{
First, we compared the analytic variational equations results with the corresponding numerical partial derivatives, also term by term.  For each component of position and velocity, we computed second order derivatives with finite differences using ``shadow particles'', pairs of real particles with their initial conditions displaced by a small amount in position 
($\epsilon\sim10^{-8}$~AU) 
and velocity ($\epsilon\sim10^{-8}$~AU/day).  Second, we compared the results from integrating the shadow particles with those from integrating the variational equations. The results verify the analytic derivatives to the precision of the numerical derivatives. }
Fig.~\ref{FIG:vary} demonstrates that results from variational equations and integrating shadow particles differ at second order in time, as expected.

\subsection{Integrations}

\begin{figure}[!tbph]
  \centering
    \includegraphics[width=\textwidth]{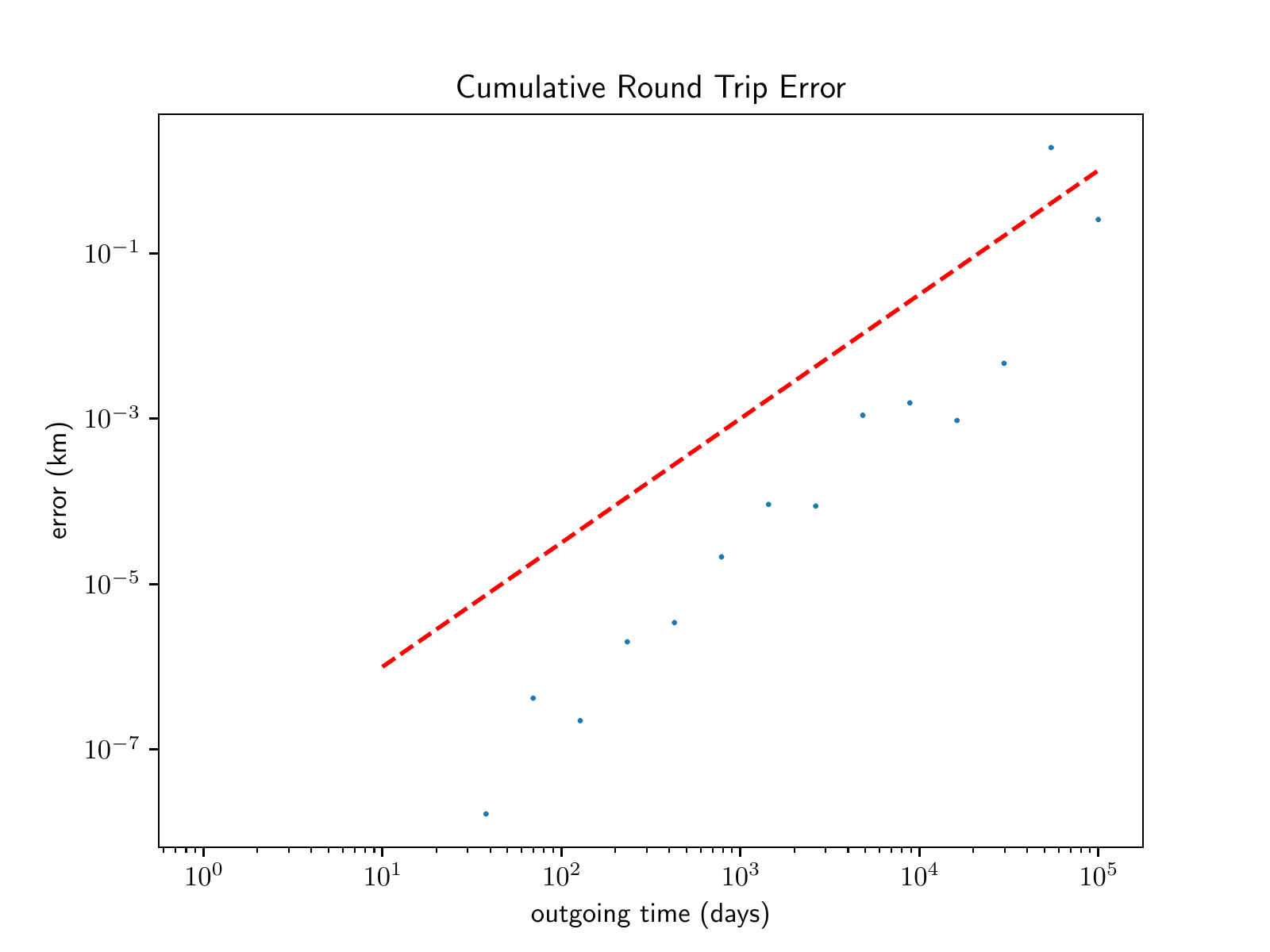}
      \caption{The magnitude of the accumulated error in integrating asteroid (3666) Holman for the indicated outgoing time and back to the initial time.  The error grows approximately as $t^{3/2}$, as is shown by the dashed red line.
      } 
  	\label{FIG:out_and_back_error}
\end{figure}

In order to verify the integration of the equations of motion, we perform a series of ``out-and-back'' tests.  These involve integrating a set of initial conditions for some period of time, reversing the direction of the integration, and integrating back to the starting time.
Because the IAS15 integrator is not intrinsically time reversible and errors accumulate in both directions, comparing the initial conditions with the final state constitutes a meaningful comparison, when dissipative terms are excluded.  Again, we use (3666) Holman for these tests.
An optimal integrator, meaning that all floating point errors are unbiased, would exhibit an energy error that grows as $\mathrm{O(n^{1/2})}$  and errors in the angles that grow as $\mathrm{O(n^{3/2})}$, 
where $n$ is the number of steps~\citep{Brouwer.1937,Rein.2015}.
The accumulated error during an out-and-back integration test is shown in Fig.~\ref{FIG:out_and_back_error}. This error grows roughly as $t^{3/2}$, indicating that any remaining biases are insignificant on this time scale. The slope may be changing toward the longer integrations; we leave investigating that to future work.

In addition to out-and-back tests, which validate the integrator against itself, we also compare the integrated results with those from JPL's small body integrator. We integrate the orbit of the main belt asteroids (3666) Holman for $10$, $10^2$, $10^3$, $10^4$, and $10^5$~days.  At the end of the integrations, we find that the positions differ by $2.7\times10^{-4}$, $9.5\times10^{-4}$, $9.3\times10^{-2}$, $3.7$, and $19$~m, respectively.  Even for the longest integration this implies an angular difference of 10~micro-arcseconds.  For (84100) Farnocchia, the differences after the same integration lengths are comparable: $1.5\times10^{-4}$, $8.5\times10^{-4}$, $1.4\times10^{-2}$, $2.4$, and $32$~m, respectively.  These integrations demonstrate that our models and integrations are essentially identical.  In addition, it demonstrates that the numerical limits of ASSIST are well below our goal of 1~milli-arcsecond over ten years.
However, it is worth noting that although the results of both routines agree the precision of the models themselves is ultimately limited by the size of the smallest terms included.

Based on these integrations, we find that ASSIST is presently comparable in speed to JPL's small body code on similar hardware.  Further speed optimization of ASSIST is left to future work.

Next we consider the trajectory of (99942) Apophis around the time of its 2029 April 13 close approach to the Earth.  This is a particularly stringent test because Apophis comes within six Earth radii from the geocenter \citep{Farnocchia2022}, which accentuates the importance of terms such as the Earth's gravitational harmonics, as well as the integration timestep.   And, perhaps more importantly, the scattering of the encounter amplifies the differences.

Starting at 2029 January 1 and integrating forward, the differences between the ASSIST and JPL small body code integrations are less than $1$~cm prior to the encounter.  The results diverge after the encounter.  However, even after integrating $10^3$~days, through the encounter and beyond, the difference is only $\sim500$~m.

\begin{figure}[!tbph]
  \centering
    \centering
  \begin{minipage}[b]{0.45\textwidth}
    \includegraphics[width=\textwidth]{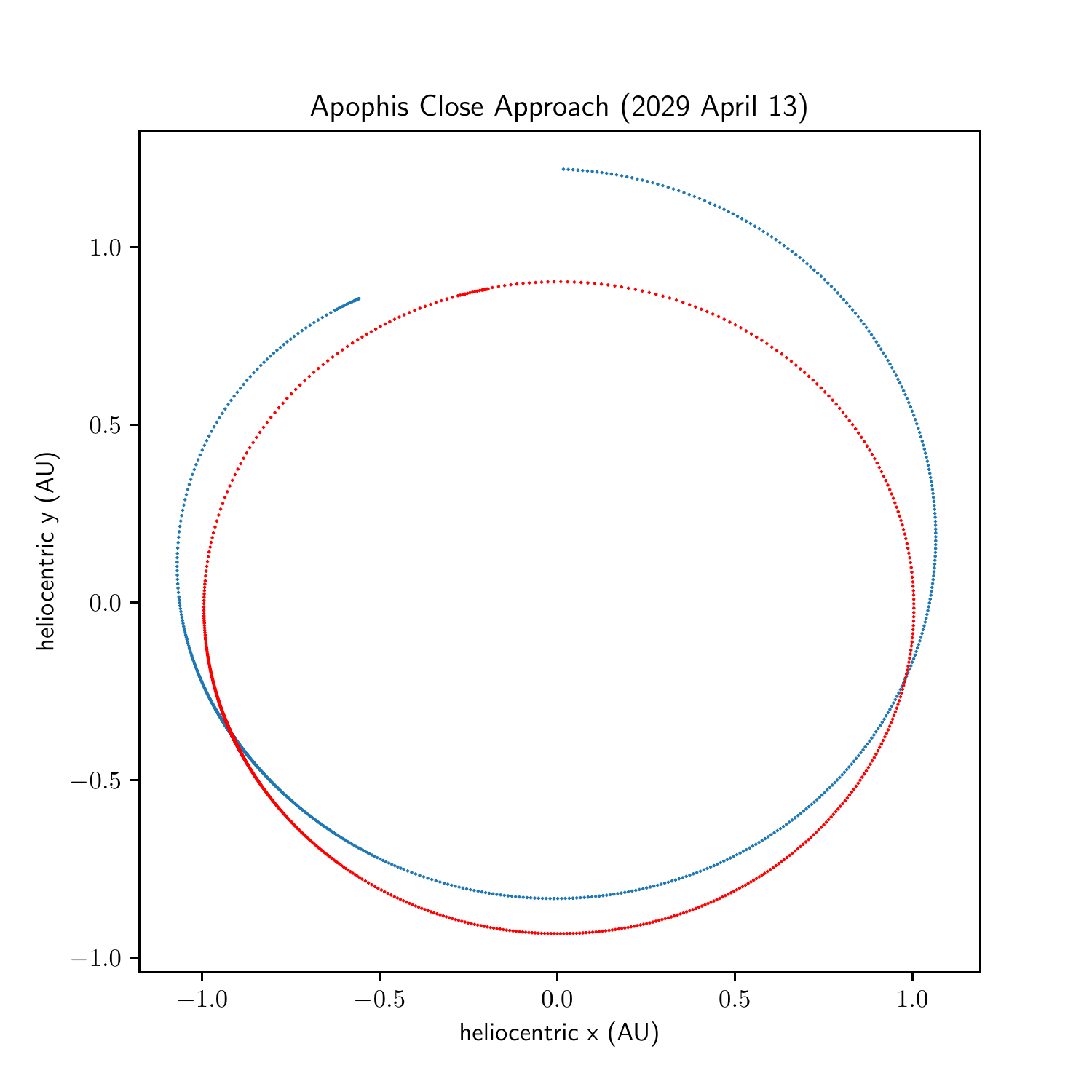}
  \end{minipage}
  \hfill
  \begin{minipage}[b]{0.45\textwidth}
    \includegraphics[width=\textwidth]{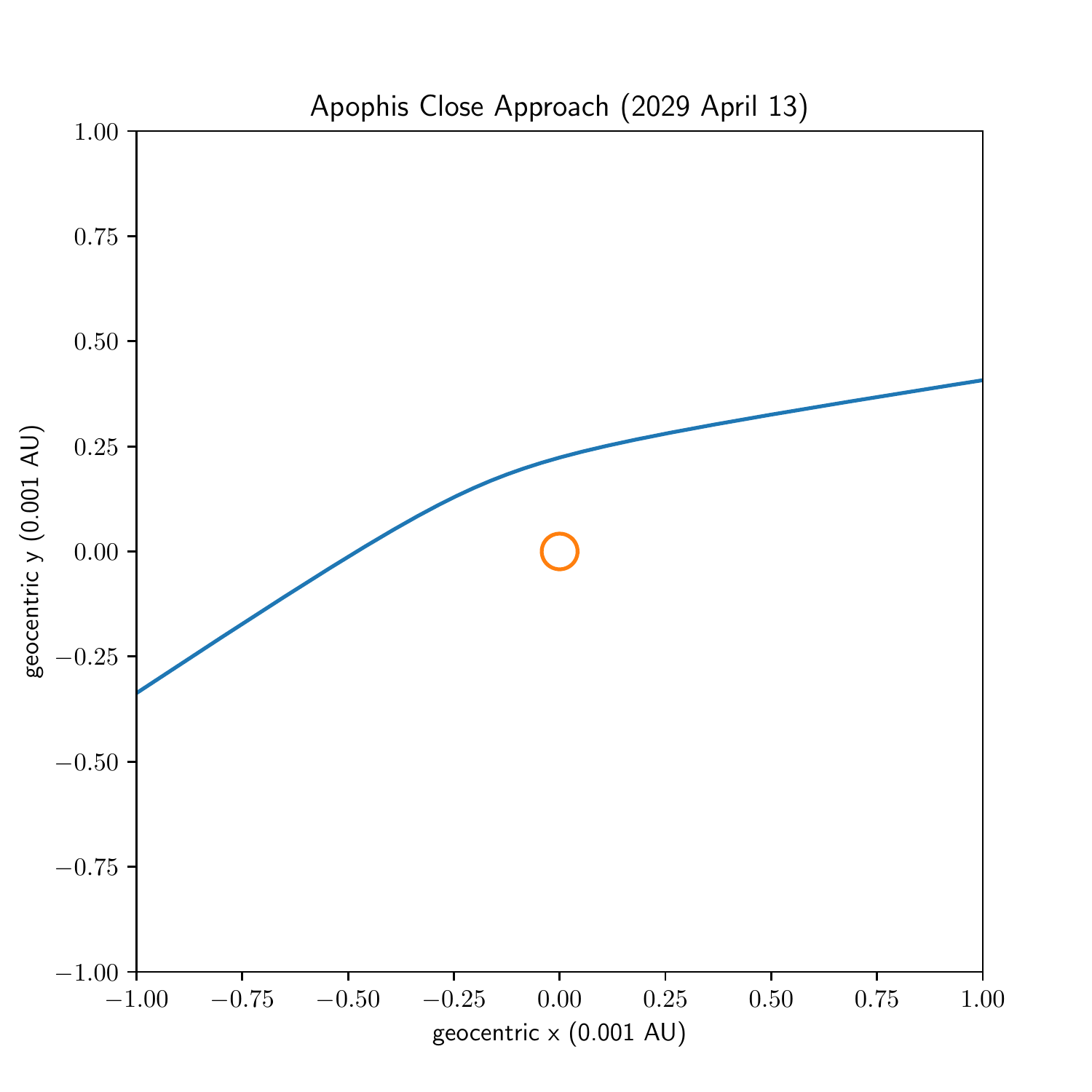}
  \end{minipage}
      \caption{Left: The heliocentric equatorial trajectories of Earth (red) and Apophis (blue) from 2029 Jan 1 to 2030 Jan 1, encompassing the close approach on 2029 April 13.  The x and y axes are ICRF coordinates.
      Right: The geocentric, equatorial position of Apophis near the time of the 2029 April 13 encounter (Earth's physical size is shown as a circle).  The deflection of the trajectory is clear.  
      }
  	\label{FIG:apophis_earth}
\end{figure}

\begin{figure}[!tbph]
  \centering
     \centering
   \begin{minipage}[b]{0.45\textwidth}     
    \includegraphics[width=\textwidth]{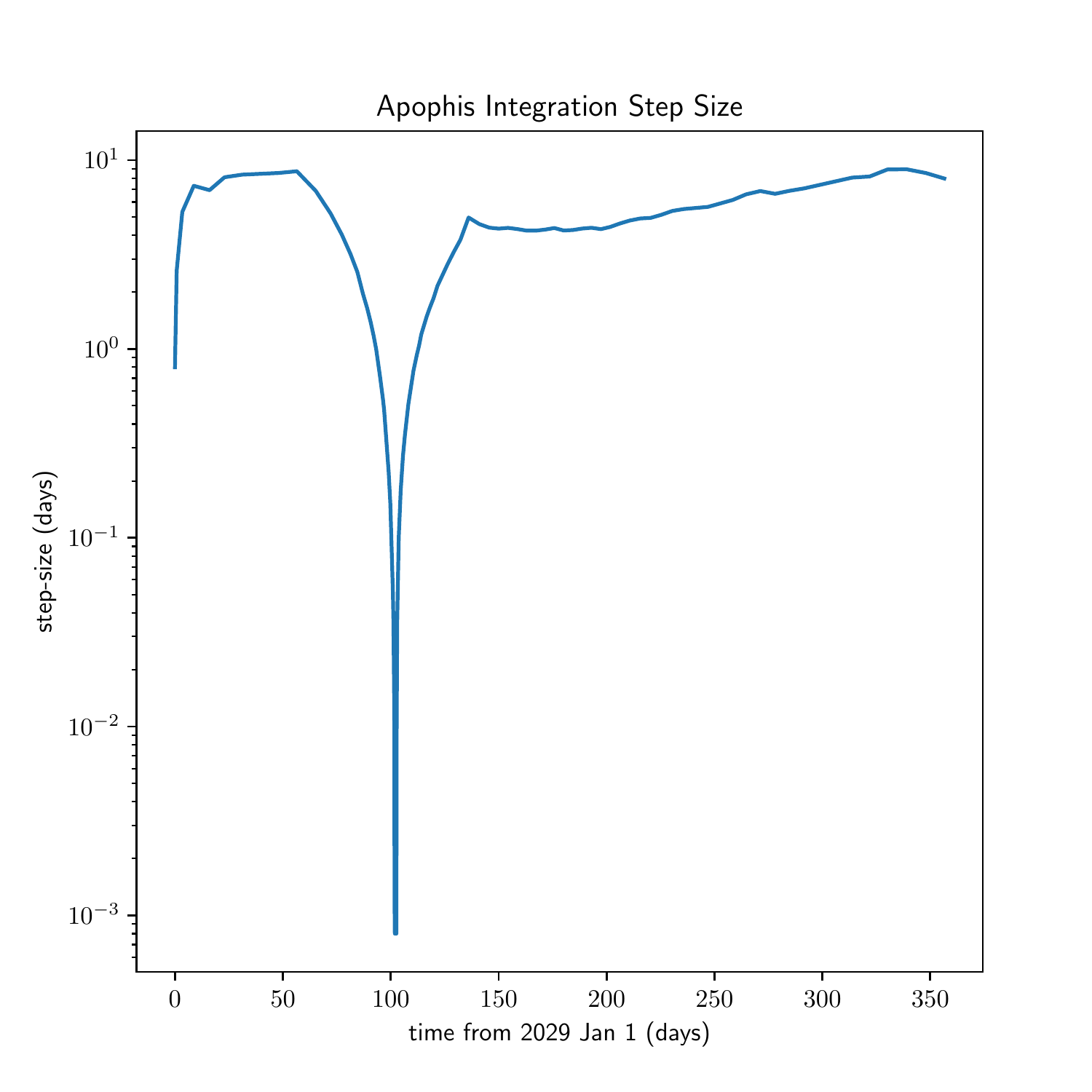}
    \end{minipage}
    \begin{minipage}[b]{0.45\textwidth}
    \includegraphics[width=\textwidth]{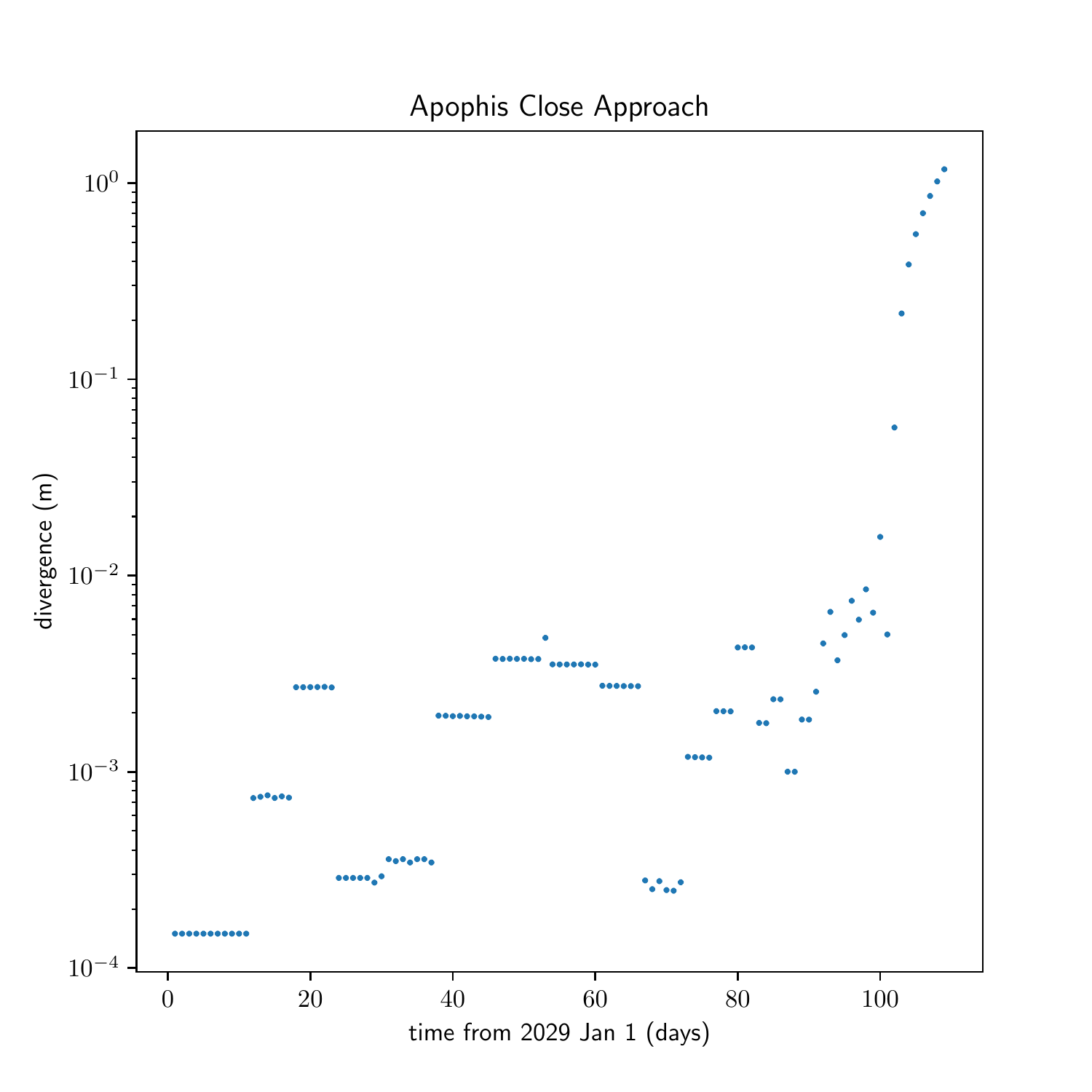}
    \end{minipage}
      \caption{Left: The IAS15 step size as a function of time from the start of the integration (2029 Jan 1).  The step size drops, reaching the prescribed minimum, at the time of the encounter.
      Right: Divergence of two calculated trajectories as a function of time, differing by a small displacement in initial conditions. Note that the time axis in the left panel extends further.} 
  	\label{FIG:apophis_integration}
\end{figure}

The heliocentric orbit of Apophis and that of the Earth, as well as the geometry of the close approach, are shown in the panels of Fig.~\ref{FIG:apophis_earth}.  The deflection of Apophis's orbit is obvious.  The integrator time step can be seen to drop significantly during close approach as indicated by the separation of the plotted points; this can be seen explicitly in left panel of Fig.~\ref{FIG:apophis_integration}, where the timestep drops to a prescribed minimum value of $10^{-3}$~days.  The divergence of our integrated orbit from the results when the initial conditions are offset in the $x$-coordinate by $10^{-15}$~AU is shown in right panel of Fig.~\ref{FIG:apophis_integration}.  This remains small despite the close encounter, indicating that the integrator is stable.

\begin{figure}[!tbph]
  \centering
    \includegraphics[width=\textwidth]{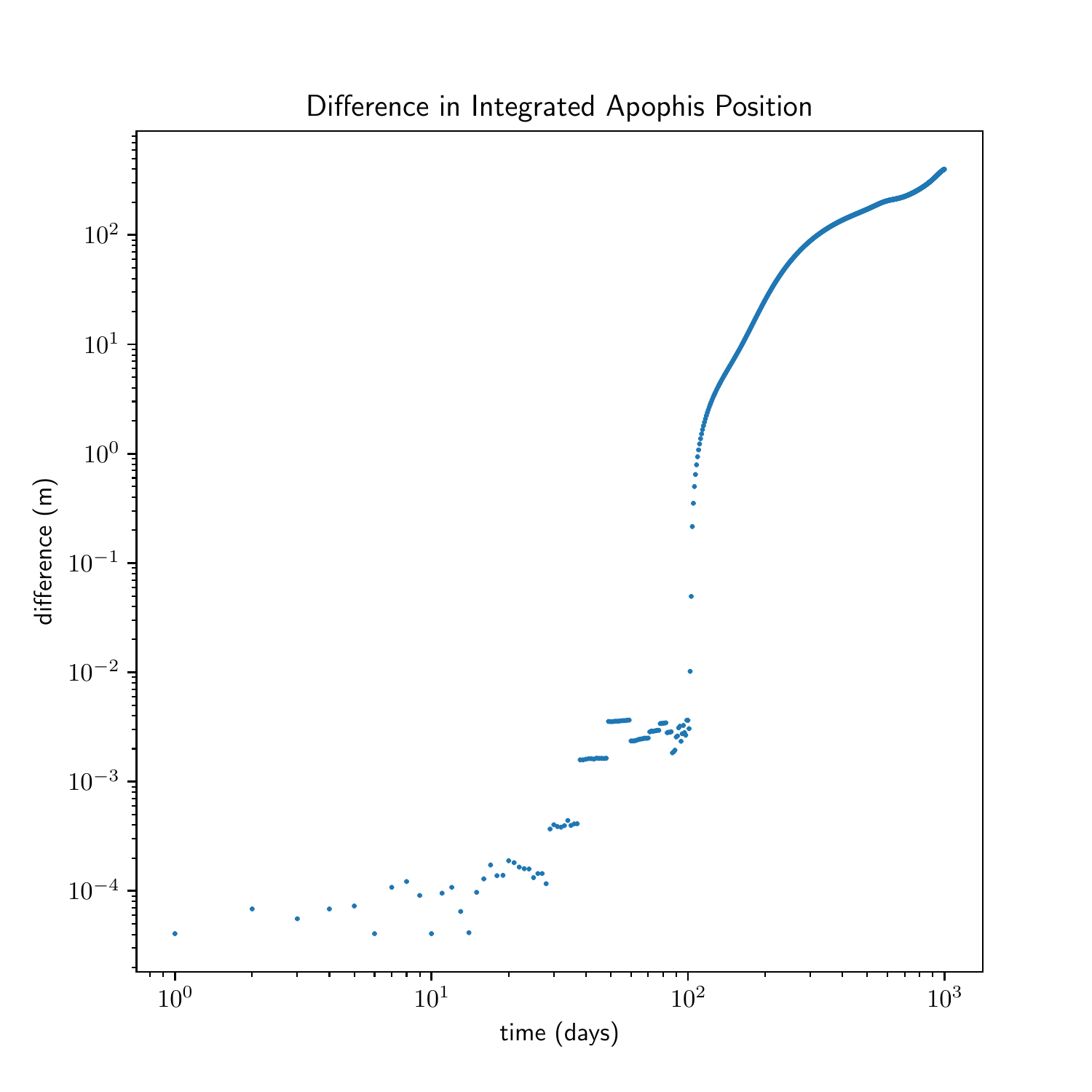}
      \caption{The magnitude of the difference in the positions of Apophis integrated with ASSIST and with the JPL small bodies code, as a function of time starting from 2029 January 1.  The differences are less than $\sim1$~cm before the encounter, near  day 102 in the figure, growing to $\sim500$~m by the end. Small differences in the calculated velocites near the close approach lead to a smoothly growing separation between the results as the integrations are continued.  The discontinuities in the difference are due to adaptive step size changes.
      } 
  	\label{FIG:apophis_comparison}
\end{figure}

Fig~\ref{FIG:apophis_comparison} shows the magnitude of the difference in the positions of Apophis integrated with ASSIST and with the JPL small bodies code, as a function of time starting from 2029 January 1.  The differences are less than $\sim1$~cm before the encounter, near  day 102 in the figure. Small differences in the calculated velocites near the close approach lead to a smoothly growing separation between the results as the integrations are continued.  The discontinuities in the difference are due to adaptive step size changes.

\begin{figure}[!tbph]
  \centering
    \includegraphics[width=\textwidth]{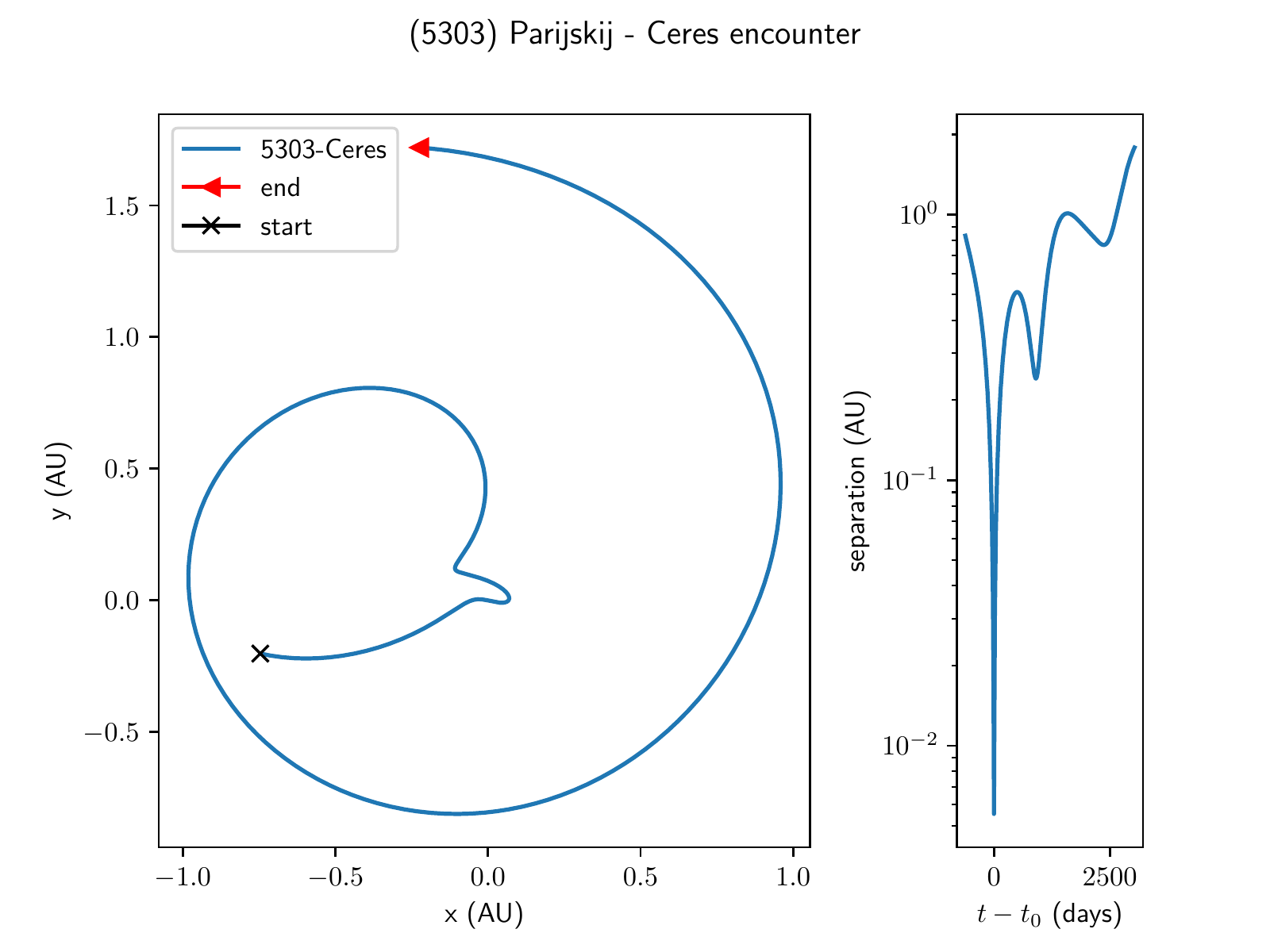}
      \caption{(5303) Parsijkij-Ceres close approach. Left: Relative position. Right: Magnitude of the separation vs time.  }
  	\label{FIG:5303}
\end{figure}

As an additional, visually interesting demonstration, we consider the serial Cereal encounters of (5303)~Parijskij,
 with closest approach 1996~September~10~\citep{Kovacevic_2007}.  The left panel of Fig.~\ref{FIG:5303} shows the xy position of 5303 relative to Ceres over a $10$~year interval starting before the closest approach.  The right panel shows the magnitude of the separation over the same time span.  The difference in the relative position differs from that of the JPL small bodies code of only $\sim1$~m.  (The difference with JPL Horizons is $\sim500$~m.)

\section{Code}
\label{SECN:CODE}

ASSIST is written in C99, following the style and structure of REBOUND.
In addition to the underlying code, we have written a Python wrapper to the basic integrator functions.
This makes ASSIST easier to use, particularly for users who are not familiar with the C99 language and associated compilers.
In addition, the Python wrapper provides a layer of abstraction upon which more complex applications can be built.

The code and the associated Python wrapper\footnote{\software{ASSIST, Holman et al (2023) : https://doi.org/10.5281/zenodo.7778017}}
 can be found at \url{https://github.com/matthewholman/assist}.
 Installation instructions can be found at \url{https://readthedocs.org/projects/assist}. 
 A limited installation for users wishing only to use the wrapper is available via the standard PyPI Python package installer, pip. 

\edit2{
 A number of examples in C99 and within Jupyter notebooks are included in the repository.  Among those is the code for reproducing the figures in this paper.
 }

\section{Conclusions}
In this paper, we have introduced ASSIST - a new, general-purpose, ephemeris quality test-particle integrator. The software is written in C99 in the REBOUND framework, and we also provide a convenient Python wrapper. The software is freely available for distribution under the GNU General Public License v3.0. Integrations are performed using the IAS15 integrator, using the positions of planets and 16 most massive asteroids from the JPL DE441 ephemeris. We find that out-and-back integration tests produce an error that grows as $n^{3/2}$, indicating optimal Brouwer's law behavior. We have verified that our results are in agreement with the JPL small-body \edit2{ integrator based on a term-by-term comparison, as well as in the integrated results for several test cases. } ASSIST achieves our goal of better than 1~milli-arcsecond precision over ten years.  We expect that ASSIST will be valuable to the solar system dynamics community as the volume of high-precision data to be fitted increases substantially with LSST and the Vera Rubin Observatory~\citep{Jones.2018,Ivezic.2019,Jones.2020}.

Future work could include extending the equations of motion to be more suitable for artificial satellite orbits, parallelizing the code to improve the speed performance, and building a framework for orbit fitting.

\label{SECN:DISCUSS}

\acknowledgments
We are grateful to the anonymous referees for their helpful suggestions.
We thank Zachary Murray for suggesting using the (5303) Parijskij - Ceres encounter as a test.  
MJH gratefully acknowledges support from the NSF (AST-2206194), the NASA YORPD Program (80NSSC22K0239), and the Smithsonian Scholarly Studies Program (2022).
DF conducted this research at the Jet Propulsion Laboratory, California Institute of Technology, under a contract with the National Aeronautics and Space Administration (80NM0018D0004).
MJP gratefully acknowledges NASA award 80NSSC22M0024.

\end{document}